# FD-DiT: Frequency Domain-Directed Diffusion Transformer for Low-Dose CT Reconstruction

Qiqing Liu, Guoquan Wei, Zekun Zhou, Yiyang Wen, Liu Shi, Qiegen Liu, Senior Member, IEEE

*Abstract*—Low-dose computed tomography (LDCT) reduces radiation exposure but suffers from image artifacts and loss of detail due to quantum and electronic noise, potentially impacting diagnostic accuracy. Transformer combined with diffusion models has been a promising approach for image generation. Nevertheless, existing methods exhibit limitations in preserving fine-grained image details. To address this issue, frequency domain-directed diffusion transformer (FD-DiT) is proposed for LDCT reconstruction. FD-DiT centers on a diffusion strategy that progressively introduces noise until the distribution statistically aligns with that of LDCT data, followed by denoising processing. Furthermore, we employ a frequency decoupling technique to concentrate noise primarily in high-frequency domain, thereby facilitating effective capture of essential anatomical structures and fine details. A hybrid denoising network is then utilized to optimize the overall data reconstruction process. To enhance the capability in recognizing high-frequency noise, we incorporate sliding sparse local attention to leverage the sparsity and locality of shallow-layer information, propagating them via skip connections for improving feature representation. Finally, we propose a learnable dynamic fusion strategy for optimal component integration. Experimental results demonstrate that at identical dose levels, LDCT images reconstructed by FD-DiT exhibit superior noise and artifact suppression compared to state-of-the-art methods.

## I. Introduction

COMPUTED tomography (CT), a widely adopted clinical imaging modality for both diagnosis and intervention, is associated with ionizing radiation exposure, raising valid concerns about increased lifetime cancer risk [1]-[3]. Reducing tube current to achieve low-dose CT (LDCT) lowers radiation exposure but increases image noise and artifacts, potentially impacting diagnostic accuracy and assessment confidence [4] [5]. Balancing radiation dose reduction with maintained image quality remains a key challenge [6][7].

LDCT reconstruction faces challenges from increased noise and loss of detail. Sinogram domain methods [8]-[10] directly operate on raw projection data, specifically designed to suppress noise and preserve details through sinogram filtering or preprocessing while taking full advantage of the original projection geometry. Concurrently, through iterative forward and back projection cycles with statistical modeling, iterative reconstruction methods [11][12] optimize both data fidelity in the sinogram domain and resultant image quality, thereby achieving comprehensive improvements in LDCT reconstruction. Alternatively, post-processing methods [13]-[15] enhance CT image quality by directly operating on the image domain, effectively balancing noise suppression and structural preservation. Building on these complementary methods, substantial research efforts have consequently been dedicated to advancing overall LDCT reconstruction performance.

Data-driven deep learning models have been widely applied to LDCT image processing in recent years. Early convolutional neural network (CNN)-based models like RED-CNN [16], performed well in denoising when mapping LDCT to normal-dose CT (NDCT) images, due to their ability to capture discriminative linear features. However, they tend to over-smooth the images with insufficient fidelity to fine details. Generative adversarial networks (GANs) have been employed to preserve realistic texture and structural information [17]-[21], but the training is inherently unstable and requires extensive optimization.

In recent years, diffusion models have attracted extensive attention due to their high-quality generation and training stability. Song *et al*. [22] explored a score-based diffusion model, achieving competitive noise reduction without compromising detail fidelity. Gao *et al*. [23] further introduced CoreDiff that enhanced LDCT reconstruction by modulating contextual error, reducing sampling steps and increasing noise sensitivity [24]. The learning mechanism of CoreDiff inspires this work. Huang *et al*. [25] proposed an unsupervised one-sample diffusion model for LDCT reconstruction, using Hankel matrix prior and iterative optimization to suppress image noise and reduce artifacts. Studies [26][27] showed that attention mechanism in transformer was capable for capturing sequential and detailed information without over-smoothing. Peebles *et al*. [28] proposed to replace the traditional U-Net with transformer as the backbone network of diffusion model to achieve high quality image generation. Recent advancements in transformer included Swin-Unet [29] and U-Vit [30], both of which utilized skip connections to improve image quality. The combination of transformer and diffusion model is worthy of exploration.

Existing LDCT reconstruction methods often underutilize information available in transform domains, potentially leading to the loss of crucial detail. Consequently, developing processing techniques that fully leverage these transform domain features is essential. For instance, Hu *et al*. [31] introduced a

This work was supported by National Natural Science Foundation of China (621220033, 62201193). (Q. Liu and G. Wei are co-first authors.) (Co-corresponding authors: L. Shi and Q. Liu.).

Q. Liu, G. Wei, Y. Wen, L Shi and Q. Liu are with School of Information Engineering, Nanchang University, Nanchang 330031, China. ({liuqiqing, guoquanwei, wenyiyang}@email.ncu.edu.cn, {shiliu, liuqiegen}@ncu.edu.cn)

Z. Zhou is with School of Mathematics and Computer Sciences, Nanchang University, Nanchang 330031, China. (zhouzekun@email.ncu.edu.cn)

domain-specific convolutional and high-frequency reconstruction network (DoCR) for image segmentation, incorporating an auxiliary high-frequency task to mitigate artifacts arising from low-frequency component substitution. Further demonstrating the value of frequency-aware processing, Xie *et al*. [32] designed a frequency-aware dynamic network for efficient single-image super-resolution, which assigned low-frequency regions to inexpensive operations and high-frequency regions to complex operations. Similarly, Mao *et al*. [33] employed fast Fourier transform [34] to flexibly capture blurring patterns, improving deblurring performance.

Based on the above findings, we present a frequency domain-directed diffusion transformer for LDCT reconstruction (FD-DiT). FD-DiT leverages multi-feature separation and a multi-network learning framework to enhance the quality of LDCT reconstruction. Multi-feature separation refers to decoupling the projection domain data into high-frequency and low-frequency components using Gaussian filtering. The multi-network learning framework comprises three distinct modules: Fourier-based high-frequency denoising module (FHD), Fourier-based low-frequency denoising module (FLD), and full-frequency denoising module (FFD). These modules are specifically designed according to the distribution of noise across different frequency domains. To optimally fuse denoised outputs from three modules, we propose learnable dynamic fusion network (LDF) that promotes distributional similarity between denoised data and ground truth data. The main contributions of this work can be summarized as follows:

- To effectively explore multi-domain information and target noise, the study employs Gaussian filtering to decouple projection domain data into high-frequency and low-frequency components, thereby enabling effective isolation of noise and genuine information during the denoising process. Simultaneously, the study introduces the self-attention mechanism of transformer to compensate for the limitations of convolutional local receptive fields. By incorporating transformer, FD-DiT achieves competitive performance in multi-feature representation perception, multi-content learning, and denoising reconstruction tasks.
- To faithfully utilize the high-frequency noise in the FHD module, the multi-head dilated attention (MHDA) mechanism incorporates sliding sparse local attention (SSLA), which adapts to the sparse and localized characteristics of shallow-layer information. Additionally, skip connections are employed to integrate these shallow features into deeper network layers, thereby enhancing feature representation.
- For optimal integration of outputs from multiple modules, LDF is proposed to learn the multi-scale information at different frequencies to capture the optimal fusion between components, and experiments prove that LDF shows the best results in reconstruction.

The remaining sections of this paper are organized as follows. Section II presents diffusion model as well as vision transformer. Section III introduces multi-network learning under the diffusion model, which expands attention mechanism and fusion strategy. Experimental results are given in Section IV, with the final conclusion drawn in Section VI. To facilitate the description of this study, we summarize the core modules of FD-DiT in Table I.

TABLE I
SUMMARY OF MODULE ABBREVIATIONS IN FD-DiT

| | Full name | | Full name |
|---|---|---|---|
| FLD | Fourier-based low-frequency denoising module | MHSA | Multi-head self-attention |
| FHD | Fourier-based high-frequency denoising module | MHDA | Multi-head dilated attention |
| FFD | Full-frequency denoising module | SSLA | Sliding sparse local attention |
| LDF | Learnable dynamic fusion module | PWLS | Penalized weighted least-squares |
| TV | Total variation | | |

## II. PRELIMINARY

### A. LDCT Sinogram Denoising

The noise in LDCT data primarily originates from quantum noise [35] and electronic noise [36]. Quantum noise is caused by statistical fluctuations in X-ray photons as they pass through the object, becoming more prominent when photon counts decrease [37]. Electronic noise, on the other hand, stems from inherent limitations in the detector and signal readout system, such as dark current noise and readout noise [38]. Under low-dose conditions, the weaker signal amplifies the impact of both types of noise, significantly reducing the signal-to-noise ratio and affecting image quality.

The denoising problem in CT sinogram data is represented by an optimization equation as follows:

$$\min_{x}\{\|x - y\|_2^2 + \mu R(x)\} \quad (1)$$

where $x$ is full-dose sinogram data, $y$ represents a low-dose sinogram data and $\|x - y\|^2$ is the data fidelity term. $R(x)$ represents the adjustment prior degree term, which is selected as the total variation (TV) semi-norm. $\|\cdot\|_2$ represents the $\ell_2$-norm. Besides, $\mu$ is a factor to keep a good balance between the data-consistency term and the regularization term.

TV [39][40] can differentiate among the countless solutions of Eq. (1) and select the optimal data with desired characteristics as the sinogram data for denoising. Typically, TV term is defined as follows:

$$R(x) = \|x\|_{TV}^2 = \int_\Omega |\nabla x| \, dx \quad (2)$$

where $\Omega$ is the bounded domain. $\nabla x$ represents the gradient of the sinogram $x$. TV term is robust in removing noise and artifacts [41].

### B. Generalized Diffusion

The generalized diffusion models build upon the foundation of standard diffusion models. It employs unique degradation mechanisms, such as blurring, masking or noise addition, in contrast to the conventional reliance on Gaussian noise. This adaptation better aligns the model with real scenarios, improving its practical utility and relevance. Specifically, given a data $x_0$, a custom degradation process $D$ is used to degrade the data from $x_0$ to $x_T$ over a total diffusion step $T$, which can be noted as $D(x_0, x_T, t)$. The denoising of LDCT data employs a degradation process analogous to the standard diffusion model, but substituting the Gaussian distribution with the empirical data distribution of LDCT data at the starting point. The definition of noise addition at any time step:

$$x_t = D(x_0, x_T, t) = \sqrt{\alpha_t} x_0 + \sqrt{(1-\alpha_t)} x_T \qquad (3)$$

where $x_T$ is the distribution of LDCT data and $\alpha_t < \alpha_{t-1}, \forall 1 \le t \le T$.

In the reverse process, we require a restoration operator $R$ to approximately reconstruct $x_0$, which can be expressed as follows:

$$R(x_t, t) \approx x_0 \qquad (4)$$

In practice, a neural network parameterized by $\theta$ is used to implement $R$. The network can be trained by solving the optimization problem defined by the following objective function:

$$\min_\theta \mathbb{E}_{x_0 \sim X} \|R_\theta(D(x_0, x_T, t) - x_0)\| \qquad (5)$$

where $\mathbb{E}_{x_0 \sim X}$ represents the expectation over all data sampled from the distribution $X$ and $\|\cdot\|$ denotes a norm, with the $\ell_1-$norm chosen in the experiment.

### C. Attention in Vision Transformer

The vision transformer (ViT) adapts transformer architecture to computer vision tasks. Multi-head self-attention (MHSA) mechanisms [42] are applied within ViT to capture global features in images. Instead of relying on local convolutional operations, MHSA utilizes the long-range dependency modeling capabilities of transformers to extract information. The global attention mechanism establishes relationships between all positions within the input through self-attention, effectively providing a global receptive field. This is achieved by computing the similarity distribution among query ($Q$), key ($K$), and value ($V$), dynamically assigning attention weights to enable each position can focus on global information. It is described as follows:

$$(Q, K, V) = Y(W_Q, W_K, W_V) \qquad (6)$$

$$Attention(Q, K, V) = Softmax\left(\frac{QK^T}{\sqrt{d_k}}\right) V \qquad (7)$$

where $Y$ is composed of patch blocks of the image, $W_Q$, $W_K$, $W_V$ mean the weight matrix of $Q, K, V$ respectively, and $d_k$ is the dimension of $K$.

Sparse local attention [43] can contribute to enhancing local attention between patches, thereby preserving fine-grained details more effectively. In this study, for a given position $Q_{(m,n)}$ in the query $Q$, a sliding window operation is applied to $K$ and $V$ to extract their local neighborhood regions, followed by a local self-attention computation with $Q_{(m,n)}$. The operation is implemented through the following equation:

$$(K, V) = Unfold(Y(W_K, W_V), k, r) \qquad (8)$$

where $Unfold(\cdot)$ represents the sliding window operation, $k$ is the size of the sliding window, $r$ is the dilation rate of the sliding window, which introduces sparsity by expanding the sampling intervals of the sliding window, thereby enabling sparse attention.

## III. METHOD

### A. Motivation

CT data can be typically decomposed into high-frequency and low-frequency components, with the high-frequency component capturing complex textures and fine details, and the low-frequency component representing the overall structural contours. By performing decomposition in Fourier domain using Gaussian filtering, as showed in Fig. 1, we observe that noise is primarily concentrated in the high-frequency band, whereas structural information exhibits a strong correlation with the low-frequency domain. Therefore, we adopt different strategies to process the decoupled high-frequency and low-frequency components.

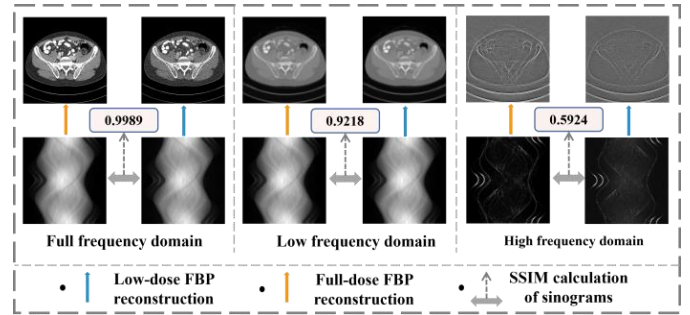

**Fig. 1.** Illustration of the contrast between low-dose and full-dose CT data in different frequency domain. It can be observed that most of the noise is distributed in the high-frequency domain.

Although the data is decomposed into high-frequency and low-frequency components, it still has limitations when denoised by U-Net. U-Net relies on convolutional operations to extract local features, which makes it difficult to capture complex and variable noise patterns when processing high-frequency data, prone to causing detail loss or noise residue. In contrast, transformer can capture long-range dependencies in data from a global perspective with its self-attention mechanism. When handling high-frequency data containing dominant noise, it can accurately identify and suppress complex noise while preserving image details. U-Net, when processing low-frequency data, efficiently smooths background noise through local feature extraction and feature fusion. Additionally, the data separation process introduces edge errors that affect the quality of sinogram data. Therefore, we innovatively adopt transformer for high-frequency data denoising, utilize U-Net for processing low-frequency data, and employ an additional U-Net for full-frequency data to reduce edge errors caused by data separation, achieving precise denoising and reconstruction of LDCT images.

### B. Overview of FD-DiT

The FD-DiT framework integrates a training process and a reconstruction process. During training process, as illustrated in Fig. 2, forward diffusion progressively transforms NDCT into LDCT data distributions via noise injection, while reverse diffusion employs a multi-network denoising architecture comprising FLD, FHD, FFD, and LDF modules. Specifically, data are decomposed into low-frequency, high-frequency, and full-frequency components through Gaussian filtering. The low-frequency and full-frequency components are processed by U-Net-based FLD and FFD modules respectively to preserve structural features and perform denoising. The full-frequency component containing dominant noise artifacts is handled by a transformer-based FHD module. LDF module subsequently synthesizes outputs from these three modules to generate denoised sinogram data, completing one step of reverse diffusion. During reconstruction process, the pre-trained denoising network executes iterative reverse diffusion enhanced by penalized weighted least-squares (PWLS) and TV regularization applied at every step. This regularization constrains the solution space, ensures data fidelity, and mitigates over-smoothing or residual noise accumulation throughout the diffusion trajectory, ultimately enabling robust recovery of high-quality CT image.

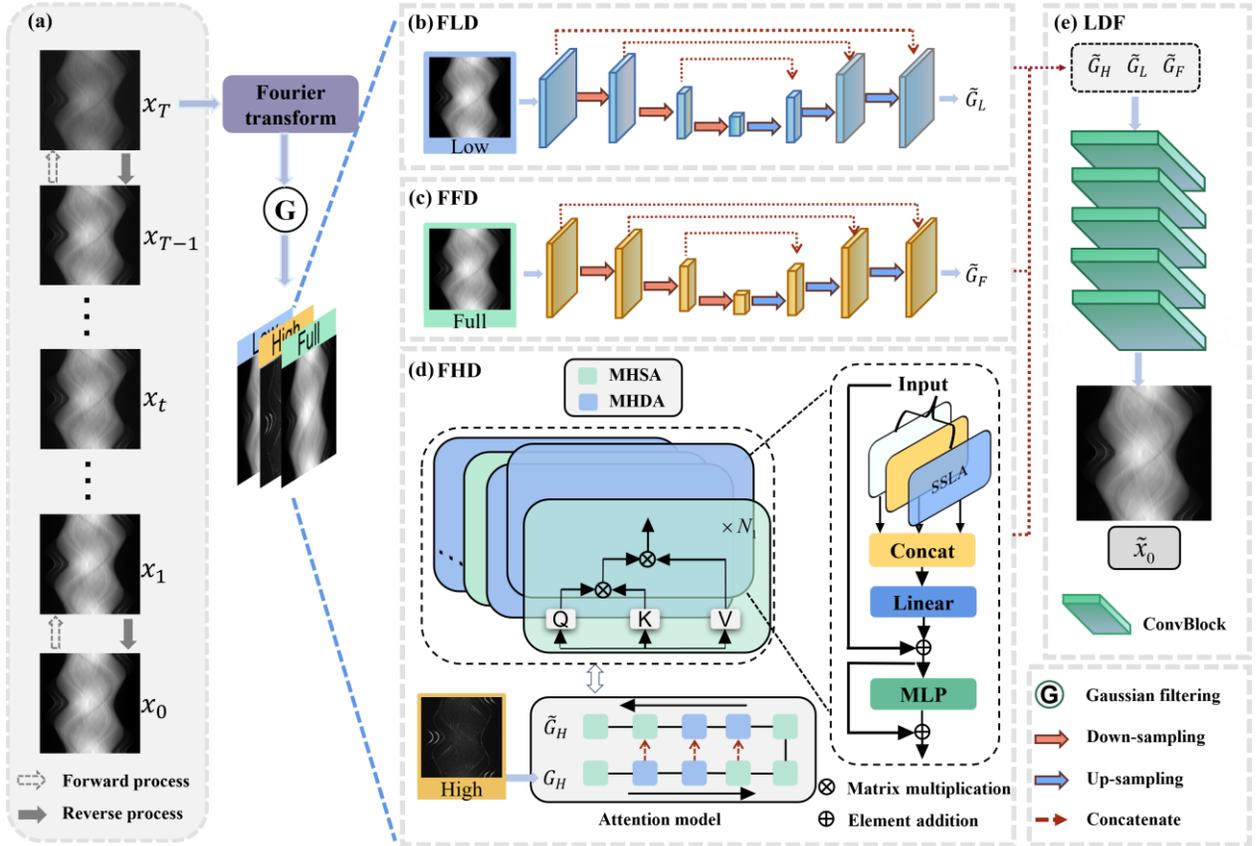

**Fig. 2.** Training process of FD-DiT. (a) Generalized diffusion model incorporating the degradation of NDCT data into LDCT data distribution, (b) mitigation of distortion caused by residual low-frequency domain noise, (c) dynamic constraints on the denoising process to balance noise suppression and structural preservation, (d) high-frequency denoising module composed of MHDA blocks and MHSA blocks, (e) learnable dynamic fusion strategy for optimal cross-frequency fusion.

### C. Training Process of FD-DiT

***Foward Diffusion:*** Unlike conventional diffusion models that perturb data toward a random Gaussian distribution, our approach employs a generalized diffusion which designates the low-dose data distribution as the terminal distribution of the forward diffusion process. The forward noise addition process of the diffusion model we used is a perturbation process introduced to maintain the mean value based on Eq. (9). It is proved to reduce CT data drift and maximize the preservation of image features during the perturbation process. The details are as follows:

$$x_t = \mathcal{P}(x_0, x_T, t) = \alpha_t x_0 + (1 - \alpha_t) x_T \qquad (9)$$

where $x_T$ is the distribution of LDCT data and $\alpha_t < \alpha_{t-1}, \forall 1 \leq t \leq T$.

***Fourier-based Reverse Diffusion:*** During each reverse diffusion step, the sinogram data undergoes Gaussian filtering to decompose into high-frequency, low-frequency, and full-frequency components. These components are processed by FLD, FHD, and FFD modules respectively, with their outputs subsequently synthesized by LDF module to obtain the denoised estimate for that diffusion step.

Based on the spectral characteristics of LDCT data noise distribution, we propose a frequency-aware decoupling strategy using Gaussian filtering in the Fourier domain. This strategy leverages the inherent property that quantum noise manifests as high-frequency random perturbations, while anatomical structures exhibit strong low-frequency correlations. The decomposition of data into low-frequency and high-frequency components via Gaussian filtering can be mathematically represented as follows:

$$\begin{cases} G_L = \mathcal{F}^{-1}(\mathcal{F}(x_t) \odot \mathcal{K}_\sigma) \\ G_H = \mathcal{F}^{-1}(\mathcal{F}(x_t) - (\mathcal{F}(x_t) \odot \mathcal{K}_\sigma)) \end{cases} \qquad (10)$$

where $\mathcal{F}$ denotes the transform in the Fourier domain and $\mathcal{F}^{-1}$ is the inverse transform. $\mathcal{K}_\sigma$ denotes the Gaussian kernel in Fourier domain with bandwidth $\sigma$, and $\odot$ represents element-wise multiplication. To streamline and expedite our explanation, we introduce shorthand notation using $G_L$, $G_H$ and $G_F$ to represent our low-frequency, high-frequency and full-frequency, respectively.

*1) High-Frequency Denoising Transformer Network:* It has been demonstrated that transformers exhibit remarkable long-range dependencies on global features [44][45]. Meanwhile, inspired by the local attention mechanisms in CNNs, they focus on modeling relationships between individual patches to mitigate the high computational costs and redundancy inherent in relying solely on global attention [46] [47]. This approach effectively addresses the inherent dichotomy between global noise distribution patterns and local structural correlations.

FHD module strategically combines global and local attention mechanisms, utilizing long-range connections to effectively capture multi-scale features within high-frequency data. Departing from conventional transformer approaches that typically partition input data into patch embeddings flattened into a single long sequence, FHD dynamically reshapes the data into a long sequence within each attention module. Furthermore, it incorporates the diffusion timestep $t$ as an additional token within this sequence. The calculation of the global attention mechanism has been presented in Eq. (6) and Eq. (7). The result of FHD can be expressed as follows:

$$\tilde{G}_H = FHD(G_H) = A_j(G_H) \qquad (11)$$

where $A_j(\cdot)$ are the attention modules of FHD, when $j = 1,4,5,6,9,10$, it represents MHSA, a global attention module, and when $j = 2,3,7,8$, it represents MHDA, a local attention module.

Recognizing the shallow-layer locality and sparsity of ViTs [48]-[50], we introduce SSLA into MHDA. Unlike traditional dynamic sparse convolution [51], which operates at the feature map level, SSLA performs self-attention at the token level. This is achieved by sparsely selecting keys and values within a sliding window centered on the query patch, as shown in Fig. 3. Formally, local attention is described as follows:

$$Attention(Q, K, V) = SSLA(Q_{(m,n)}, K_r, V_r, r)$$
$$= Softmax\left(\frac{Q_{(m,n)} K_r^T}{\sqrt{d_k}}\right) V_r \qquad (12)$$

where $Q_{(m,n)}$ represents the position $(m,n)$ in $Q$. $K_r$ and $V_r$ are the keys and values that selected from the feature maps $K$ and $V$ with the dilation rates $r =$ 1, 2, 3. SSLA sparsely selects keys and values with a sliding window of size $\omega \times \omega$ centered at $(m,n)$ to perform self-attention. It can use $(m', n')$ for local self-attention:

$$\{(m', n') \mid m' = m + p \times r, n' = n + q \times r\}$$
$$\frac{\omega}{2} \le p, q \le \frac{\omega}{2} \qquad (13)$$

*2) Low-frequency and Full-frequency Denoising U-Net:* Recognizing that low-frequency components often retain subtle yet critical structural information despite minor noise contamination, we preserve low-frequency features to enhance reconstruction fidelity. To mitigate distortion caused by residual noise in these components, we introduce FLD. This module selectively refines low-frequency representations while preserving key details through adaptive frequency-domain filtering.

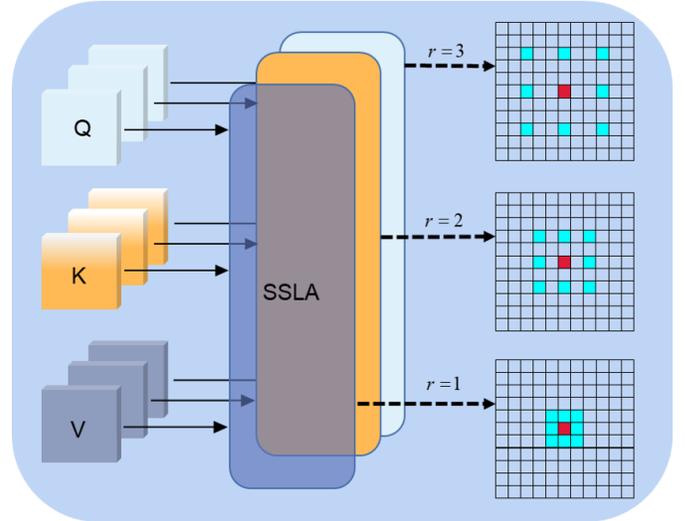

**Fig. 3.** Illustration of sliding sparse local attention (SSLA). SSLA decomposes queries, keys, and values into multi-channel structures, performing local convolution operations within sparse blocks (highlighted in red) around query patches, with distinct dilation rates for diversified receptive fields. By default, the dilation rates are set to $r$ = 1, 2, 3, corresponding to receptive field sizes of $3 \times 3$, $5 \times 5$, and $7 \times 7$ for different SSLA modules.

To address instability from high-frequency and low-frequency decoupling effects, we deploy a dual-branch U-Net architecture, FLD and FFD. The low-frequency branch processes noise-separated features through FLD, while FFD maintains an identical U-Net structure to handle raw data features. FFD branch acts as a compensation mechanism, dynamically constraining denoising to balance noise suppression with structural preservation, particularly when frequency boundaries are ambiguous. FLD and FFD can be described as:

$$\tilde{G}_L = FLD(G_L) \qquad (14)$$
$$\tilde{G}_F = FFD(G_F) \qquad (15)$$

where $\tilde{G}_L$ represents the low-frequency projection domain data reconstructed after FLD, and $\tilde{G}_F$ represents the projection domain data reconstructed after FFD containing all the features of low and high frequencies, which is used as a constraint or compensation factor.

*3) Learnable Dynamic Fusion Network (LDF):* Although the collaborative framework successfully addresses the decoupling task, optimally integrating the three pre-optimized components remains a significant challenge. To resolve this, we propose LDF composed of multi-channel convolutional modules. This strategy adaptively learns interactions between multi-scale features and dynamically optimizes the fusion effectiveness of frequency-domain components. Traditional Fourier transform-based fusion methods rely on fixed weighting coefficients, neglect interdependencies across frequencies, and consequently suffer from suboptimal feature fusion. In contrast, LDF overcomes these limitations through adaptive learnable mechanisms, achieving superior feature fusion. The presentation is described as follows:

$$X = Concat(\tilde{G}_H, \tilde{G}_L, \tilde{G}_F) \qquad (16)$$

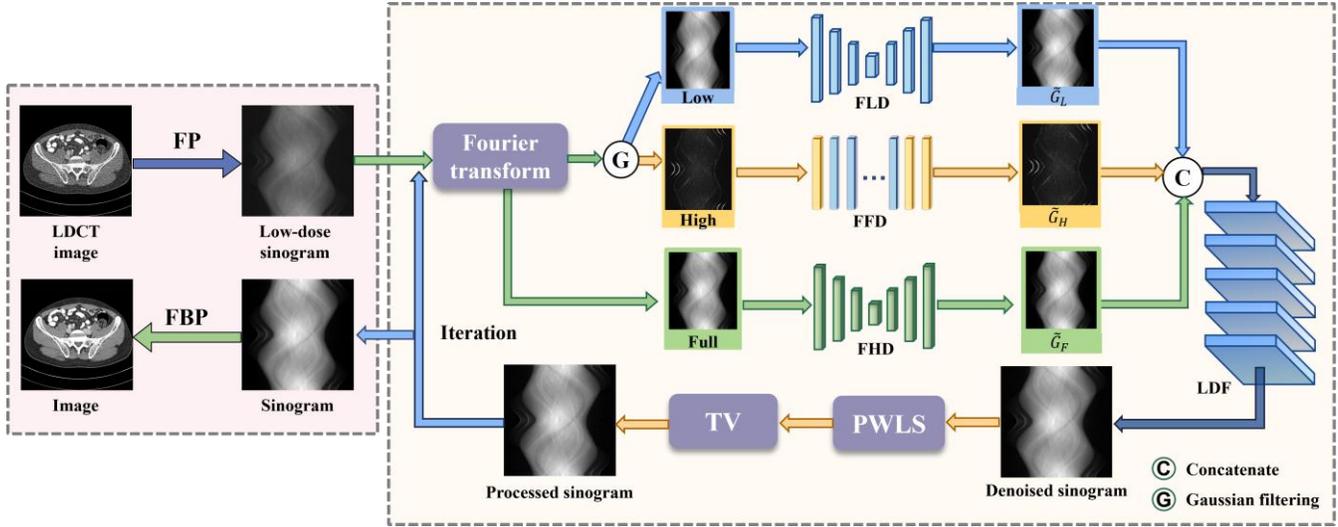

**Fig. 4.** The pipeline for iterative reconstruction stage of FD-DiT. During the reconstruction process, LDCT reconstruction is performed by processing different frequency domains in parallel through multiple modules.

$$\tilde{x}_0 = LDF(X) \quad (17)$$

*4) Network Parameters Optimization:* During the training process, we obtain the neural network recovery operator $\mathcal{N}$ parameterized by $\theta$. Here, the detailed process of $\mathcal{N}$ is what we present in subsections 1, 2 and 3 of this chapter. It can be described as follows:

$$\tilde{x}_0 = \mathcal{N}_\theta(x_t, t) \quad (18)$$

The objective optimization function to be achieved during the training process is defined as follows:

$$\min_\theta \mathbb{E}_{x_0 \sim \emptyset, x_T \sim \varphi} \|\mathcal{N}_\theta(\mathcal{P}(x_0, x_T, t), t) - x_0\|^2 \quad (19)$$

where $\emptyset$ is the NCDT data distribution and $\varphi$ represents the LDCT data distribution.

### D. Reconstruction Process of FD-DiT

The reconstruction process of FD-DiT is demonstrated in Fig. 4. At first, the sinogram data $x_T$ is transformed from the LDCT image $I$ via operation $F^{-1}$ which corresponds to forward projection (FP):

$$x_T = F^{-1}(I) \quad (20)$$

Subsequently, $x_T$ is processed by the trained network $\mathcal{N}_\theta$. It is worth noting that in the iterative reconstruction, the data is decomposed into $G_L(x_t)$, $G_H(x_t)$ and $G_F(x_t)$ through Eq. (10) before entering the denoising network $\mathcal{N}_\theta$. The resulting output from $\mathcal{N}_\theta$ can be described as follows:

$$\begin{cases} \tilde{G}_H(x_t) = FHD(G_H(x_t)) \\ \tilde{G}_L(x_t) = FLD(G_L(x_t)) \\ \tilde{G}_F(x_t) = FFD(G_F(x_t)) \end{cases} \quad (21)$$

$$x_t = LDF(Concat(\tilde{G}_H(x_t), \tilde{G}_L(x_t), \tilde{G}_F(x_t))) \quad (22)$$

Furthermore, we incorporate PWLS and TV as a fidelity term in the iterative reconstruction process. It prevents both excessive smoothing from individual denoising steps and the accumulation of residual noise due to insufficient denoising.

It uses PWLS to obtain the best evaluation of the noisy sinogram. The standard PWLS can be described as follows:

$$x = \arg\min_x[(x - y)^T W(x - y) + \mu R(x)] \quad (23)$$

**Algorithm: FD-DiT**

**Training Stage**

**Dataset:** Paired NDCT/LDCT data set $x_0, x_T$, steps $T$

1: **Repeat**
2:     Calculate $x_t = \mathcal{P}(x_0, x_T, t), t \sim ([1, T])$
3:     $\tilde{x}_0 = \mathcal{N}_\theta(x_t, t)$
4:     **Take a gradient descent step on:**
       $\nabla_\theta \|\mathcal{N}_\theta(\mathcal{P}(x_0, x_T, t), t) - x_0\|^2$
5: **Until** converged
6: **Trained** FD-DiT

**Reconstruction Stage**

**Dataset:** Initial data $x_T = F^{-1}(I)$ (**FP**)

1: **Load** the trained $\mathcal{N}_\theta$
2: **For** $t = T$ to 1 **do**
3:     Decoupling $x_t$ as $G_L, G_H$ by Eq. (10)
4:     $\tilde{x}_t \leftarrow \mathcal{N}_\theta(x_t, t)$:
       Calculate $\tilde{G}_H = FHD(G_H)$
       Calculate $\tilde{G}_L = FLD(G_L)$
       Calculate $\tilde{G}_F = FFD(G_F)$
       Calculate $\tilde{x}_t = LDF(Concat(\tilde{G}_H, \tilde{G}_L, \tilde{G}_F))$
5:     $x_{t-1} = \frac{W(y - \tilde{x}_t) + \mu R'(\tilde{x}_t)}{W + \mu}$ (**PWLS**)
6:     $x_{t-1} = TV(x_{t-1})$ (**TV**)
7: **End for**
8: **Final** image $\tilde{I} = F(x_0)$ (**FBP**)
9: **Return** $\tilde{I}$

where superscript T represents the transposing operation. Eq. (23) can be further solved as:

$$x_{t-1} = \frac{W(y - x_t) + \mu R'(x_t)}{W + \mu} \quad (24)$$

To decrease the influence of noise, the scale coefficient $\eta$ for system calibration is set to 22000.

$$W = diag\{w_t\} = diag\left\{\frac{1}{\sigma_{x_t}^2}\right\} \quad (25)$$
$$= diag\left\{\frac{1}{l_1}e^{\frac{\eta}{x_t}}\right\}$$

TV minimization is introduced as a statistical denoising method in the sinogram domain to obtain optimal reconstructed data for our denoising task. It is employed to remove both noise and artifacts, serving as a critical component in the iterative reconstruction process. Specifically, TV minimization can be stated as follows:

$$x_{t-1} = x_t - \alpha \parallel x - x_{t-1} \parallel \frac{\nabla x_{t-1}}{\parallel \nabla x_{t-1} \parallel^2} \quad (26)$$

where $\alpha$ is the length of each gradient-descent step.

Once the reconstructed projection $x_0$ is obtained, the final image $\tilde{I}$ can be obtained:

$$\tilde{I} = F(x_0) \quad (27)$$

where $F(\cdot)$ means the filtered back-projection (FBP).

## IV. EXPERIMENTS

### A. Data Specification

**AAPM Challenge Dataset:** This study employed the Mayo Clinic LDCT Grand Challenge dataset to evaluate the performance of LDCT imaging algorithms. The dataset consisted of simulated human abdominal CT images provided by the Mayo Clinic, including full-dose CT scans from 10 patients. Data from 9 patients, totaling 4742 slices, were used for training, and data from 1 patient were reserved for evaluation. Each slice had a resolution of $512 \times 512$ pixels and a thickness of 1 mm.

The artifact-free images reconstructed from the normal-dose projection data using the FBP algorithm were regarded as the ground truth. To simulate the quantum noise induced by low-dose conditions, we extracted noisy projection data with photon counts of $1 \times 10^5$, $5 \times 10^4$, and $1 \times 10^4$ from normal-dose projections. For fan-beam CT reconstruction, the Siddon's ray-driven algorithm [52] was utilized to generate the projection data. The distance from the rotation center to the source and detector was set to 40 cm. The detector width was 41.3 cm, composed of 720 detector elements. Over 360 projection views were uniformly distributed throughout the entire imaging process.

**CIRS Phantom Dataset:** To validate the generalization capability of the FD-DiT, we conducted additional experiments using an anthropomorphic CIRS phantom scanned on a GE Discovery HD750 CT system. For fairness, the model pre-trained on the AAPM Challenge dataset was directly applied to the CIRS dataset across multiple dose levels without fine-tuning. Test results were averaged over 21 independent test samples to ensure statistical robustness.

### B. Model Training and Parameter Selection

This experiment trained FD-DiT model using the PyTorch framework on two NVIDIA GeForce RTX 4090D GPUs with 24 GB. The Adam optimizer was employed with an initial learning rate of $2 \times 10^{-4}$ for 300k training iterations. For the transformer architecture, it primarily consisted of 2 modular blocks, including MHDA modules with depths of 1 and 2, and an MHSA module with a depth of 1. In PWLS scheme, the number of iterations was set to 22,000, while TV minimization used 2 iterations. The source code can be publicly accessed at: https://github.com/yqx7150/FD-DiT.

### C. Reconstruction Experiments

**AAPM Reconstruction Results:** To assess the efficacy of the proposed algorithm, we conducted comprehensive comparisons with several state-of-the-art methods, including FBP [1], RED-CNN [16], CoreDiff [23], DU-GAN [21], U-ViT [30], RAP [53], and WiTUnet [54]. All competing methods were trained and tested using their original hyperparameters to ensure fair evaluation. CT reconstruction was performed on simulated data at three dose levels (1e4, 5e4, and 1e5 photons).

Table II summarizes the quantitative results of PSNR, SSIM, and MSE on the AAPM Challenge dataset. For reconstructed images across different projection views, FD-DiT achieves the highest PSNR, SSIM, and MSE values, with optimal metrics highlighted in bold. Notably, FD-DiT demonstrates significant improvements over existing approaches when benchmarked against the dataset's ground-truth images, underscoring its superiority in noise suppression and detail preservation.

Fig. 5 and Fig. 6 illustrate the reconstruction results of different dose levels with 1e4, 5e4, and 1e5 photons respectively. Visual analysis highlights the superiority of FD-DiT in restoring edge contours while effectively suppressing noise and artifacts. In the zoomed ROI and residual maps, RED-CNN removes noise but fails to recover holistic structures. DU-GAN struggles with sparse feature restoration. CoreDiff approximates NDCT quality but lacks edge precision. RAP and U-ViT exhibit residual artifacts despite advanced regularization, such as Hankel matrices and PWLS. WiTUnet balances texture prediction but blurs fine details. In contrast, FD-DiT's residual maps reveal near-perfect retention of micron-level anatomical features, such as tissue interfaces, and robust suppression of both stochastic and structured noise. Quantitative metrics further confirm its dose-agnostic performance in Table II, achieving the highest PSNR/SSIM and lowest MSE across all dose levels, with particularly notable gains at 1e4 photons level through a 1.2 dB PSNR improvement over U-ViT.

**CIRS Phantom Reconstruction Results:** To thoroughly validate the robustness of our learning scheme, we trained the model on the AAPM Challenge dataset and evaluated its reconstruction performance on the CIRS Phantom test data. We compared FD-DiT against baseline methods—FBP, RED-CNN, CoreDiff, DU-GAN, U-ViT, RAP, and WiTUnet with 1e4, 5e4, and 1e5 photons. The means PSNR, SSIM, and MSE metrics for the CIRS Phantom are summarized in Table II, with the best values highlighted in bold.

Obviously, compared with other methods, FD-DiT method has obvious quantitative index advantages. Table II and Fig. 7 present the quantitative results and reconstructed image features obtained from the simulated CIRS dataset, respectively. Notably, FD-DiT outperforms all competing approaches, dem-onstrating superior quantitative metrics. It is worth emphasizing that the conventional FBP exhibits markedly inferior performance on the CIRS modeling dataset with a maximum difference of 14.92 dB compared to our method and the blue-magnified regions of interest and residual maps clearly reveal its poor noise suppression. The CNN-based RED-CNN delivers stable results with minor metric fluctuations, but it still lags significantly behind our approach, sacrificing texture details in the noise removal process.

TABLE II
RECONSTRUCTION PSNR/SSIM/MSE ($10^{-4}$) OF AAPM CHALLENGE DATA AND CIRS PHANTOM DATA USING DIFFERENT METHODS AT DIFFERENT DOSE LEVELS WITH 1e4, 5e4, AND 1e5 PHOTONS

| Method | AAPM Challenge dataset | | | CIRS phantom dataset | | |
|---|---|---|---|---|---|---|
| | 1e4 | 5e4 | 1e5 | 1e4 | 5e4 | 1e5 |
| FBP | 25.19/0.5978/31.0 | 29.83/0.7347/10.7 | 33.62/0.8427/4.83 | 20.55/0.4048/88.5 | 28.36/0.7240/14.7 | 32.34/0.8343/5.89 |
| RED-CNN | 34.43/0.8998/3.95 | 40.93/0.9723/1.92 | 41.28/0.9742/1.29 | 32.04/0.9014/7.59 | 37.33/0.9721/2.94 | 37.51/0.9534/2.54 |
| WiTUnet | 37.11/0.9429/2.38 | 40.71/0.9696/0.91 | 42.27/0.9762/0.63 | 34.54/0.9394/3.55 | 38.19/0.9707/1.54 | 39.61/**0.9781**/1.14 |
| DU-GAN | 35.78/0.9291/3.33 | 39.09/0.9537/1.31 | 40.02/0.9638/1.06 | 30.08/0.8703/9.89 | 36.82/0.9566/2.13 | 37.48/0.9606/1.81 |
| U-ViT | 34.77/0.8758/3.46 | 36.23/0.8828/2.40 | 36.95/0.8881/2.03 | 33.08/0.8753/4.93 | 34.55/0.8762/3.52 | 35.32/0.8862/2.94 |
| RAP | 37.01/0.9014/2.04 | 41.07/0.9539/0.80 | 42.45/0.9667/0.60 | 35.22/0.9348/3.07 | 39.20/0.9611/1.22 | 40.87/0.9715/0.84 |
| CoreDiff | 39.20/0.9421/1.26 | 41.06/0.9726/0.85 | 43.98/0.9806/0.43 | 31.03/0.9121/8.04 | 35.18/0.9723/3.14 | 37.33/0.9695/1.94 |
| FD-DiT | **39.76/0.9543/1.06** | **43.08/0.9753/0.50** | **44.61/0.9811/0.35** | **35.47/0.9498/2.92** | **39.92/0.9786/1.04** | **41.29**/0.9758/**0.76** |

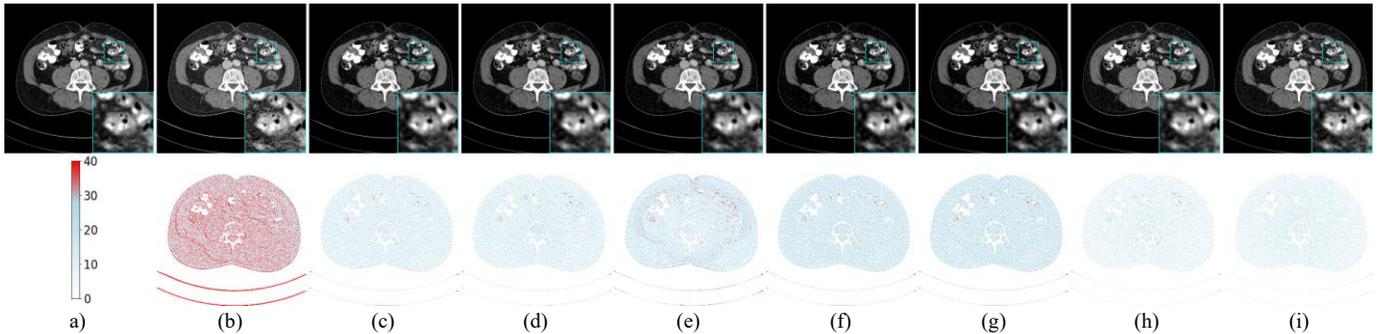

Fig. 5. Reconstruction results of AAPM Challenge data with 1e5 photons using different methods. (a) The reference image, (b) FBP, (c) RED-CNN, (d) WiTUnet, (e) DU-GAN, (f) U-Vit, (g) RAP, (h) CoreDiff, and (i) FD-DiT. The display window is [-5, 300] HU. The zoomed-in blue ROIs allow clear observation of the noise removal effect.

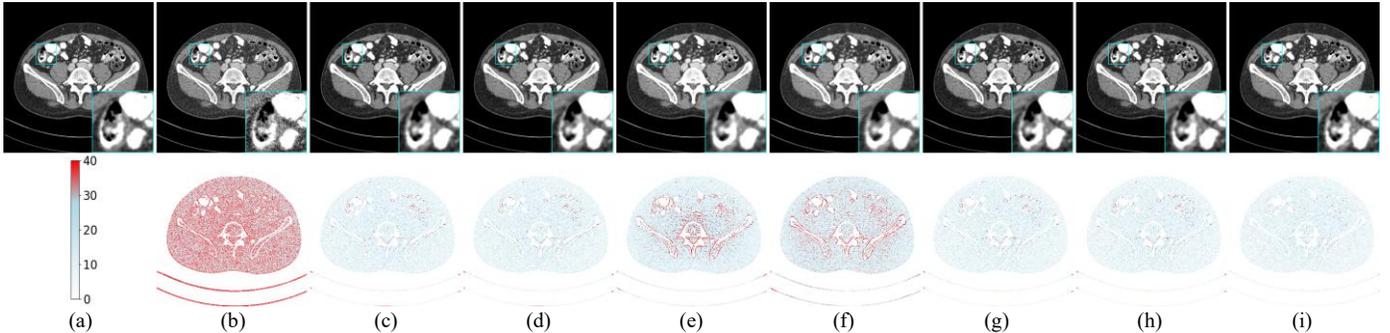

Fig. 6. Reconstruction results of AAPM Challenge data with 5e4 photons using different methods. (a) The reference image, (b) FBP, (c) RED-CNN, (d) WiTUnet, (e) DU-GAN, (f) U-Vit, (g) RAP, (h) CoreDiff, and (i) FD-DiT. The display window is [-5, 300] HU. The zoomed-in blue ROIs allow clear observation of the noise removal effect.

Furthermore, both CoreDiff and DU-GAN display generalization capabilities that are substantially inferior to their performance on the AAPM Challenge dataset, with reconstruction quality deteriorating as dose levels decrease.

Similarly, the performance of U-ViT is unremarkable, as it suffers from pronounced noise and artifacts. In contrast, the results from RAP and WiTUnet are close to those of FD-DiT, further validating the advantages of U-Net and transformer architectures. Overall, our proposed approach demonstrates significant performance improvements across various metrics and evaluation criteria, effectively addressing noise suppression, texture preservation, and detail recovery. Specifically, it consistently provides higher accuracy, better retention of fine details, and clearer image reconstructions.

### D. Comparative Experiments in FD-DiT

To fully illustrate the superiority of FD-DiT design for denoising task, we design the comparative experiments in the aspects of ablation study, attention mechanism and fusion method. The performance of the experimental design components across varying training iterations is demonstrated in Fig. 8.

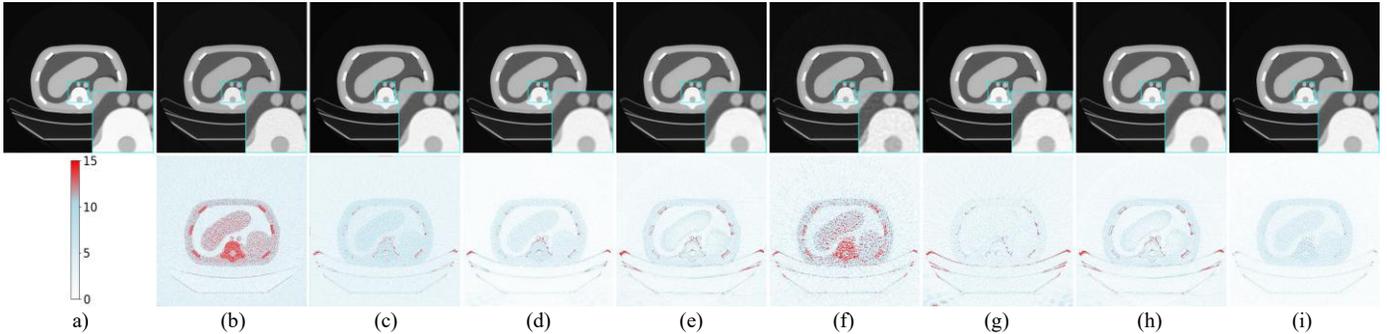

Fig. 7. Reconstruction results of CIRS phantom data with 1e5 photons using different methods. (a) The reference image, (b) FBP, (c) RED-CNN, (d) WiTUnet, (e) DU-GAN, (f) U-Vit, (g) RAP, (h) CoreDiff, and (i) FD-DiT. The display window is [-5, 300] HU. The zoomed-in blue ROIs allow clear observation of the noise removal effect.

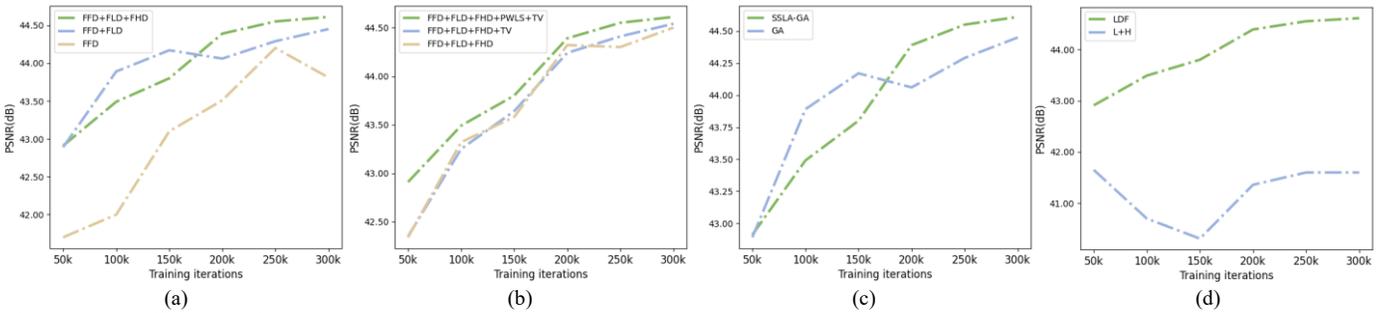

Fig. 8. Training efficiencies of distinct components. (a) FFD, FFD+FLD, and FFD+FLD+FHD, (b) PWLS+TV, (c) SSLA-GA represents the combination of MHDA and MHSA modules, GA consists of only MHSA blocks, (d) LDF denotes the fusion of high-frequency and low-frequency through a module, L+H represents the summation of high-frequency and low-frequency.

*1) Ablation Study:* FD-DiT frequency-domain module consists of FLD, FHD, and FFD. In this study, we gradually remove modules as ablation experiments to verify the effectiveness of each component. The results of the experiments are shown in Fig. 9 and Table III.

These results validate that modules (FLD, FHD) enhance the denoising capability of FD-DiT, supporting our improvements in each step are reasonable and effective for the denoising task.

*2) Sliding Spare Local Attention (SSLA) vs Only Global Attention (GA):* We replace the process involving SSLA with the process of global attention mechanism. To ensure the introduction of the SSLA does not bring excessive redundancy to features and can simultaneously reduce the cost in noise removal, we abbreviate it as SSLA-GA and GA. The proposed SSLA is demonstrated in Fig. 10 and Table IV to outperform the full global attention mechanism.

*3) Fusion Methods:* Comparative experiments are made in Fourier domain high-frequency and low-frequency summation fusion (abbreviated as L+H) and learnable dynamic fusion approach (LDF).

Through the meticulous examination of Fig. 11 and Table IV, we can derive the conclusion that the learnable dynamic fusion approach is capable of learning the mutual information fusion of different frequency domain components. This approach effectively circumvents the feature loss associated with the summation method and overcomes the limitations inherent in fixed-weight fusion.

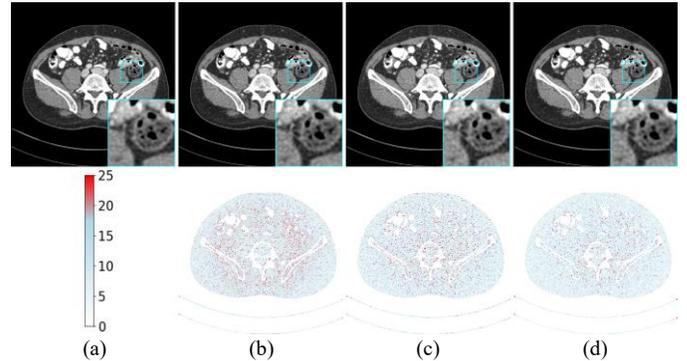

Fig. 9. Reconstruction results with 1e5 photons using different structural compositions. (a) The reference image, (b) FFD, (c) FFD-FLD, and (d) FFD-FLD-FHD. The display window in the first row is still [-5, 300] HU.

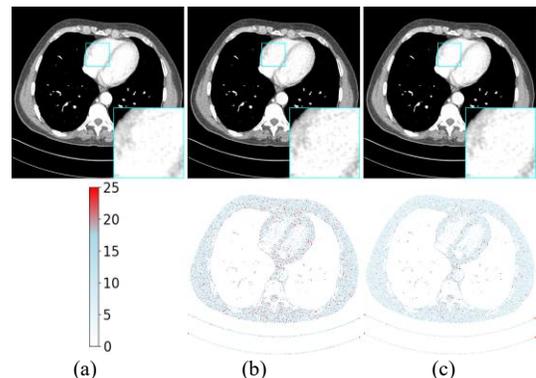

Fig. 10. Reconstruction results with 1e5 photons using different attention mechanisms. (a) The reference image, (b) GA, global attention mechanism. (c) SSLA-GA, integrated global-local attention mechanism. The display window in the first row is still [-5, 300] HU.

TABLE III
QUANTITATIVE PSNR/SSIM/MSE($10^{-4}$) RESULTS ON STRUCTURAL COMPOSITION FOR AAPM CHALLENGE DATA WITH 1E5 PHOTONS

| FFD | FLD | FHD | PSNR(dB) | SSIM | MSE |
|---|---|---|---|---|---|
| √ | - | - | 43.86 | 0.9779 | 0.42 |
| √ | √ | - | 44.27 | 0.9787 | 0.38 |
| √ | √ | √ | **44.61** | **0.9801** | **0.35** |

TABLE IV
QUANTITATIVE PSNR/SSIM/MSE($10^{-4}$) RESULTS ON ATTENTION MECHANISM AND FUSION METHOD FOR AAPM CHALLENGE DATA WITH 1E5 PHOTONS

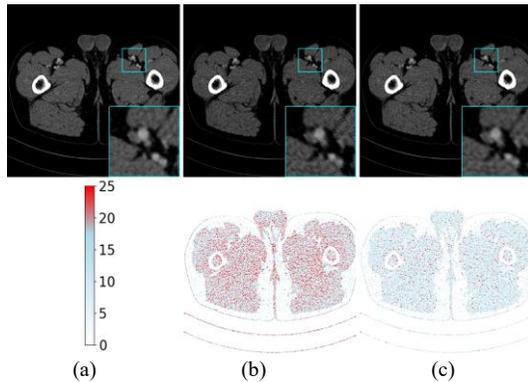

(a) (b) (c)

**Fig. 11.** Reconstruction results with 1e5 photons using different fusion methods. (a) The reference image, (b) L+H, the summation of high-frequency and low-frequency, (b) LDF. The display window in the first row is still [-5, 300] HU.

## V. DISCUSSION

FD-DiT preserves original CT features by diffusing NDCT toward the LDCT data distribution. The reverse diffusion process circumvents the need for extensive iterative sampling to reconstruct NDCT, substantially lowering computational overhead. Additionally, the high-frequency low-frequency separation strategy is conducive to learning the noise distribution of LDCT data more effectively. FD-DiT's three frequency-domain denoising modules (FHD, FLD, FFD) synergistically combine U-Net and transformer architectures, where transformer's global attention mechanism mitigates the locality constraints of U-Net's convolutions, enabling multi-perspective denoising of LDCT data. While the FD-DiT demonstrates promising results, the lack of down-sampling operations between transformer blocks may hinder the full exploitation of the multi-scale potential of dilated convolutions. Going forward, we plan to combine U-Net-like hierarchical down-sampling with dilated attention to construct a multi-scale feature pyramid.

## VI. CONCLUSION

In this study, we proposed frequency domain-directed diffusion transformer for LDCT reconstruction (FD-DiT), a new method for removing the noise presenting in sinogram domain. This method innovatively decoupled the high-frequency and low-frequency information to form a triple-frequency with the full-frequency information, providing more feature information to enhance the reconstruction capability of the model. In addition, the transformer architecture was introduced to complement the limitation of the traditional diffusion model on the global sensory field. Also, SSLA was introduced to adapt to the localization and sparsity of the shallow layer, and the long connection was adopted to transfer the shallow information to the deep layer to enhance the characterization ability of the high-frequency information. Finally, to optimize the utilization of denoised data, a learnable dynamic fusion strategy was proposed to handle the multi-network output. In both AAPM dataset and CIRS dataset, FD-DiT showed strong generalization ability, and its robustness and effectiveness were experimentally verified.

## REFERENCES

[1] D. J. Brenner and E. J. Hall, "Computed tomography—An increasing source of radiation exposure," *New Engl. J. Med.*, vol. 357, no. 22, pp. 2277-2284, 2007.

[2] J. C. Wen, V. Sai, B. R. Straatsma, and T. A. McCannel, "Radiation-related cancer risk associated with surveillance imaging for metastasis from choroidal melanoma," *JAMA ophthalmology*, vol. 131, no. 1, pp. 58–63, 2013.

[3] National Research Council et al. "Health risks from exposure to low levels of ionizing radiation," *National Academies Press*, Washington, DC, 2006.

[4] J. Wang et al., "Bismuth shielding, organ-based tube current modulation, and global reduction of tube current for dose reduction to the eye at head CT," *Radiology*, vol. 35, no. 2, pp. 170–175, 2012.

[5] M. K. Kalra et al., "Strategies for CT radiation dose optimization," *Radiology.*, vol. 230, no. 3, pp. 1000-1008, 2004.

[6] C. Catalano, M. Francone, A. Ascarelli, M. Mangia, I. Iacucci, and R. Passariello, "Optimizing radiation dose and image quality," *European Radiology Supplements*, vol. 17, pp. 26–32, 2007.

[7] L. W. Goldman, "Principles of CT: radiation dose and image quality," *Journal of Nuclear Medicine Technology*, vol. 35, no. 4, pp. 213–225, 2007.

[8] J. Wang, H. Lu, T. Li, and Z. Liang, "Sinogram noise reduction for low-dose CT by statistics-based nonlinear filters," in *Medical Imaging 2005: Image Processing*, vol. 5747, SPIE, 2005, pp. 2058–2066.

[9] T. Li, X. Li, J. Wang, J. Wen, H. Lu, J. Hsieh, and Z. Liang, "Nonlinear sinogram smoothing for low-dose X-ray CT," *IEEE Trans. Nucl. Sci.*, vol. 51, no. 5, pp. 2505–2513, 2004.

[10] P. J. La Rivière, J. Bian, and P. A. Vargas, "Penalized-likelihood sinogram restoration for computed tomography," *IEEE Trans. Med. Imaging,* vol. 25, no. 8, pp. 1022–1036, 2006.

[11] M. Beister, D. Kolditz, and W. A. Kalender, "Iterative reconstruction methods in X-ray CT," *Physica Medica*, vol. 28, no. 2, pp. 94–108, 2012.

[12] Y. Yamada et al., "Model-based iterative reconstruction technique for ultralow-dose computed tomography of the lung: a pilot study," *Investigative Radiology*, vol. 47, no. 8, pp. 482–489, 2012.

[13] Y. Chen et al., "Artifact suppressed dictionary learning for low-dose CT image processing," *IEEE Trans. Med. Imaging*, vol. 33, no. 12, pp. 2271–2292, 2014.

[14] X. Y. Cui, Z. G. Gui, Q. Zhang, H. Shangguan, and A. H. Wang, "Learning-based artifact removal via image decomposition for low-dose CT image processing," *IEEE Trans. Nucl. Sci*, vol. 63, no. 3, pp. 1860–1873, 2016.

[15] Y. Chen et al., "Thoracic low-dose CT image processing using an artifact suppressed large-scale nonlocal means," *Physics in Medicine & Biology*, vol. 57, no. 9, p. 2667, 2012.

[16] H. Chen, Y. Zhang, M. K. Kalra, F. Lin, Y. Chen, P. Liao, J. Zhou, and G. Wang, "Low-dose CT with a residual encoder-decoder convolutional neural network," *IEEE Trans. Med. Imaging.*, vol. 36, pp. 2524–2535, 2017.

[17] X. Wang and A. Gupta, "Generative image modeling using style and structure adversarial networks," in *Proc. Eur. Conf. Comput. Vis. (ECCV)*, Springer, 2016, pp. 318–335.

[18] J. Liu, Y. Zou, and D. Yang, "SemanticGAN: Generative adversarial networks for semantic image to photo-realistic image translation," in *Proc. IEEE Int. Conf. Acoust., Speech Signal Process. (ICASSP)*, 2020, pp. 2528–2532.

[19] S. S. T. Moghadam et al., "WBT-GAN: Wavelet based generative adversarial network for texture synthesis," in *Proc. 2021 11th Int. Conf. Comput. Eng. Knowl. (ICCKE)*, 2021, pp. 441–446.

[20] Q. Yang et al., "Low-dose CT image denoising using a generative adversarial network with Wasserstein distance and perceptual loss," *IEEE Trans. Med. Imaging*, vol. 37, pp. 1348–1357, 2018.

[21] Z. Huang, J. Zhang, Y. Zhang, and H. Shan, "DU-GAN: Generative adversarial networks with dual-domain U-Net based discriminators for low-dose CT denoising," *IEEE Trans. Instrum. Meas.*, vol. 71, pp. 1–12, 2021.

[22] Y. Song, J. Sohl-Dickstein, D. P. Kingma, A. Kumar, S. Ermon, and B. Poole, "Score-based generative modeling through stochastic differential equations," in *Proc. Int. Conf. Learn. Represent.*, 2020.

[23] Q. Gao et al., "CoreDiff: Contextual error-modulated generalized diffusion model for low-dose CT denoising and generalization," *IEEE Trans. Med. Imaging.*, vol. 43, no. 2, pp. 411–423, 2024.

[24] Q. Gao and H. Shan, "CoCoDiff: A contextual conditional diffusion model for low-dose CT image denoising," in *Developments in X-Ray Tomography XIV*, vol. 12242, SPIE, 2022, pp. 92–98.

[25] B. Huang, S. Lu, L. Zhang, B. Lin, W. Wu, and Q. Liu, "One-sample diffusion modeling in projection domain for low-dose CT imaging," *IEEE Trans. Radiat. Plasma Med. Sci*, 2024.

[26] T. Ye, L. Dong, Y. Xia, Y. Sun, Y. Zhu, G. Huang, and F. Wei, "Differential transformer," *arXiv:2410.05258*, 2024.

[27] J. Su, M. Ahmed, Y. Lu, S. Pan, W. Bo, and Y. Liu, "Roformer: Enhanced transformer with rotary position embedding," *Neurocomputing*, vol. 568, p. 127063, 2024.

[28] W. Peebles and S. Xie, "Scalable diffusion models with transformers," in *Proc. IEEE/CVF Int. Conf. Comput. Vis.*, 2023, pp. 4195–4205.

[29] H. Cao, Y. Wang, J. Chen, D. Jiang, X. Zhang, Q. Tian, and M. Wang, "Swin-Unet: Unet-like pure transformer for medical image segmentation," in *Proc. Eur. Conf. Comput. Vis. (ECCV)*, 2022, pp. 205–222.

[30] F. Bao, S. Nie, K. Xue, Y. Cao, C. Li, H. Su, and J. Zhu, "All are worth words: A vit backbone for diffusion models," in *Proc. IEEE/CVF Conf. Comput. Vision Pattern Recognit.*, 2023, pp. 22669–22679.

[31] S. Hu, Z. Liao, and Y. Xia, "Domain specific convolution and high frequency reconstruction based unsupervised domain adaptation for medical image segmentation," in *Proc. Int. Conf. Med. Image Comput. Comput.-Assist. Intervent.*, 2022, pp. 650–659.

[32] W. Xie, D. Song, C. Xu, C. Xu, H. Zhang, and Y. Wang, "Learning frequency-aware dynamic network for efficient super-resolution," in *Proc. IEEE/CVF Int. Conf. Comput. Vis. (ICCV)*, 2021, pp. 4288–4297.

[33] X. Mao, Y. Liu, F. Liu, Q. Li, W. Shen, and Y. Wang, "Intriguing findings of frequency selection for image deblurring," in *Proc. AAAI Conf. Artif. Intell. (AAAI)*, vol. 37, no. 2, 2023, pp. 1905–1913.

[34] L. Chi, B. Jiang, and Y. Mu, "Fast Fourier convolution," in *Proc. Adv. Neural Inf. Process. Syst.*, vol. 33, 2020, pp. 4479–4488.

[35] J. Ma, Z. Liang, Y. Fan, Y. Liu, J. Huang, W. Chen, H. Lu, "Variance analysis of X-ray CT sinograms in the presence of electronic noise background," *Med. Phys.*, vol. 39, no. 7, pp. 4051-4065, 2012.

[36] Q. Xie, D. Zeng, Q. Zhao, D. Meng, Z. Xu, Z. Liang, J. Ma, "Robust low-dose CT sinogram preprocessing via exploiting noise-generating mechanism," *IEEE Trans. Med. Imaging*, vol. 36, no. 12, pp. 2487-2498, 2017.

[37] J. Wang, T. Li, H. Lu, and Z. Liang, "Penalized weighted least-squares approach to sinogram noise reduction and image reconstruction for low-dose X-ray computed tomography," *IEEE Trans. Med. Imaging*, vol. 25, no. 10, pp. 1272–1283, 2006.

[38] Y. Fan, A. Zamyatin, S. Nakanishi, "Noise simulation for low-dose computed tomography," in *Proc. IEEE Nucl. Sci. Symp. Med. Imag. Conf. (NSS/MIC)*, 2012, pp. 3641–3643.

[39] Y. Zhang, W. Zhang, Y. Lei, and Zhou J, "Few-view image reconstruction with fractional-order total variation," *J. Opt. Soc. Am.*, vol. 31, no. 5, pp. 981-995, 2014.

[40] S. V. M. Sagheer, and S. N. George, "Denoising of low-dose CT images via low-rank tensor modeling and total variation regularization," *Artif. Intel. in Med.*, vol. 94, pp. 1-17, 2019.

[41] X. Jia, Y. Lou, R. Li, W. Y. Song and S. B. Jiang, "GPU-based fast cone beam CT reconstruction from undersampled and noisy projection data via total variation," *Med. Phys.*, vol. 37, no. 4, pp. 1757-1760, 2010.

[42] J. Devlin, M.-W. Chang, K. Lee, and K. Toutanova, "Bert: Pre-training of deep bidirectional transformers for language understanding," in *Proc. Conf. North Amer. Chapter Assoc. Comput. Linguistics (NAACL)*, 2019, pp. 4171-4186.

[43] J. Jiao, Y. M. Tang, K. Y. Lin, Y. Ding, W. Wu, and J. Yan, "Dilateformer: Multi-scale dilated transformer for visual recognition," *IEEE Trans. Multimedia.*, vol. 25, pp. 8906–8919, 2023

[44] A. Dosovitskiy, et al, "An image is worth 16×16 words: transformers for image recognition at scale," in *Proc. Int. Conf. Learn. Represent. (ICLR)*, 2020.

[45] H. Touvron, M. Cord, M. Douze, F. Massa, A. Sablayrolles, and H. Jégou, "Training data-efficient image transformers & distillation through attention," in *Proc. Int. Conf. Mach. Learn. (ICML)*, PMLR, 2021, pp. 10 347–10 357.

[46] X. Chu, Z. Tian, Y. Wang, B. Zhang, H. Ren, X. Wei, H. Xia, and C. Shen, "Twins: Revisiting the design of spatial attention in vision transformers," in *Proc. Annu. Conf. Neural Inf. Process. Syst. (NeurIPS)*, 2021, vol. 34, pp. 9355–93661.

[47] A. Hassani, S. Walton, J. Li, S. Li, and H. Shi, "Neighborhood attention transformer," in *Proc. IEEE/CVF Conf. Comput. Vis. Pattern Recognit (CVPR)*, 2023, pp. 6185–6194.

[48] Z. Liu, Y. Lin, Y. Cao, H. Hu, Y. Wei, Z. Zhang, S. Lin, and B. Guo, "Swin transformer: Hierarchical vision transformer using shifted windows," in *Proc. IEEE/CVF Int. Conf. Comput. Vis. (ICCV)*, 2021, pp. 10 012–10 022.

[49] Z. Tu, H. Talebi, H. Zhang, F. Yang, B. Behnam, Y. Li, J. Luo, and R. Ba, "MaxVit: Multi-axis vision transformer," in *Proc. Eur. Conf. Comput.Vis. (ECCV)*, Springer, 2022, pp. 459–479.

[50] W. Wang, L. Yao, L. Chen, X. Shao, J. Shi, Z. Gao, and A. Zhang, "Crossformer: A versatile vision transformer based on cross-scale attention," *IEEE Trans. Pattern Anal. Mach. Intell.*, vol. 46, no. 5, pp. 3123-3136, 2023.

[51] Y. Chen, X. Dai, M. Liu, D. Chen, L. Yuan, and Z. Liu, "Dynamic convolution: Attention over convolution kernels," in *Proc. IEEE/CVF Conf. Comput. Vis. Pattern Recognit. (CVPR)*, 2020, pp. 11030–11039.

[52] R. L. Siddon, "Fast calculation of the exact radiological path fora three-dimensional CT array," *Med. Phys.*, vol. 12, no. 2, pp. 252–255, 1985.

[53] W. Zhang, B. Huang, S. Chen, X. Li, and Y. Wu, "Low-rank angular prior guided multi-diffusion model for few-shot low-dose CT reconstruction," *IEEE Trans. Comput. Imaging*, 2024.

[54] B. Wang, F. Deng, P. Jiang, X. Zhang, and L. Liu, "WiTUnet: A U-shaped architecture integrating CNN and transformer for improved feature alignment and local information fusion," *Sci. Rep.*, vol. 14, no. 1, p. 25525, 2024.